\documentclass[twocolumn,prl,showpacs,preprintnumbers]{revtex4}
\usepackage{amssymb}
\usepackage{nicefrac}
\usepackage{graphicx}
\usepackage{epstopdf}
\usepackage{dcolumn}
\usepackage{bm} 
\usepackage[squaren]{SIunits} 
\usepackage{hyperref} 
\usepackage{upgreek}

\newcommand{\RF}{\mbox{$\text{\tiny{RF}}$}}
\newcommand{\RFC}{\mbox{$\text{\tiny{RF0}}$}}

\newcommand{\TA}{\mbox{$\text{\tiny{TA}}$}}

\newcommand{\LA}{\mbox{$\text{\tiny{L}}$}}
\newcommand{\gF}{\mbox{$g_{\text{\tiny{F}}}$}}

\newcommand{\mF}{\mbox{$m_{\text{\tiny{F}}}$}}
\newcommand{\muB}{\mbox{$\mu_{\text{\tiny{B}}}$}}

\begin{document}

\title{Time-Averaged Adiabatic Potentials:\\
Versatile traps and waveguides for ultracold quantum gases }

\author{I. Lesanovsky}{}
\author{W. von Klitzing}{}
\affiliation{Institute of Electronic Structure and Laser, Foundation
for Research and Technology - Hellas, P.O. Box 1527, GR-71110
Heraklion, Greece}

\date{\today}

\begin{abstract}
We demonstrate a novel class of trapping potentials, time-averaged
adiabatic potentials (TAAP) which allows the generation of a large
variety of traps and waveguides for ultracold atoms. Multiple traps can be coupled
through controllable tunneling barriers or merged altogether. We
present analytical expressions for pancake-, cigar-, and ring- shaped
traps. The ring-geometry is of particular interest for guided
matter-wave interferometry as it provides a perfectly smooth
waveguide of controllable diameter, and thus a tunable sensitivity of the
interferometer.
\end{abstract}


\maketitle
One of the cornerstones in the exploration of quantum-degenerate
gases has been the steady improvement in trapping techniques for neutral
atoms. The time-orbiting potentials (TOP) \cite{Petrich1995PRL},
Ioffe-Pritchard (IP) \cite{Pritchard1983PRL} and clover-leaf  traps
\cite{Mewes1996PRL} made possible the first experiments with
Bose-Einstein Condensation (BEC) of weakly interacting atomic gases
\cite{Andrews1996S,Anderson1995BEC}. Later, dipole traps allowed
spin independent trapping \cite{Stamper-Kurn1998PRL}, and the first
periodic trapping potentials where created using optical lattices
\cite{Morsch2001PRL}. Microchip traps now offer the possibility to
design complex coil configurations in a small space
\cite{Folman2002AAMOP, Fortagh2006RMP}. Radio frequency (RF) induced adiabatic potentials allow the creation of pancake-shaped traps with extreme aspect ratios \cite{Zobay2001PRL}. However, the main
challenge remains the creation of flexible, complex trap shapes
that preserve the high degree of coherence of Bose-Einstein
Condensates.  In this letter we present a novel type of trap:
time-averaged adiabatic potentials (TAAP). They allow the
creation of a great variety of trap shapes, e.g.\ stacks of pancakes, rows of
cigars or waveguides, and multiple rings or half moons. The traps can be coupled and  merged. Moreover, the trap shapes can be adiabatically transformed from one into another. TAAP structures can be created far away from their field generating coils and therefore do not exhibit any small-scale corrugation which might destroy the coherence of a matter-wave traveling inside them.

We start by considering atoms in a magnetic trap subject to a modulation in the confining field.
 The behavior of these atoms with respect to the frequency of the modulation falls  into three different regimes: the quasi-static, the time-averaging, and the RF-dressing regime. In the quasi-static regime the modulation is so slow that the atoms follow adiabatically. In the time-averaging regime the atoms are confined to a trap which is the time-average of the modulated potential. The modulation frequency $\omega_{\text{m}}$ has to be much larger than the trap frequency $\omega$ so that the atoms do not follow the modulation.  An example of a time-averaged potential is the TOP trap \cite{Petrich1995PRL,Tiecke2003JOB}. In the RF-dressing regime the modulation creates an RF field which resonantly couples the atomic Zeeman levels of the static magnetic trap. These coupled states are described by dressed atomic potentials. For the coupling to occur the frequency $\omega_{\RF}$ of the RF field has to be close to the Larmor frequency $\Omega_{\LA} =\left|\gF \muB \mathbf{B}(\mathbf{r})\right|/\hbar\,$ that is associated with the difference in energy between the Zeeman levels. The $\gF$ is the Land\'e $g$-factor of the considered hyperfine manifold and $\muB$ is the Bohr magneton.

\begin{figure}[htbp]
\includegraphics[angle=0,width=8cm]{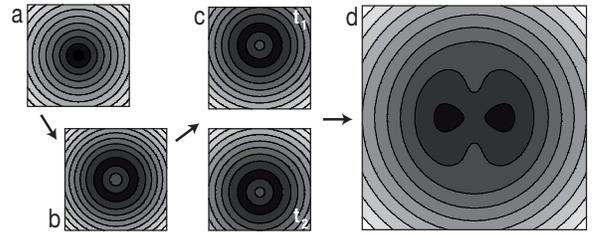}
\caption{Schematic generation of a TAAP trap; density plots of the potential energy of the atoms for different elements of the TAAP. The darker shades represent smaller values of the potential. Starting with a static magnetic trapping potential (a) we apply an RF-field resulting in the dressed potential (b). We now add a modulation field thus periodically shifting the trap up (t$_{1}$) and down (t$_{2}$) as shown in (c).  We do this at such a high frequency that the atoms are not  able to follow the modulation of the dressed potential of (c). This results in the TAAP (d).  }
\label{fig:TAAP}
\end{figure}

The TAAP combines the RF-dressing and the time-averaging regimes in order to produce trapping geometries of high complexity and flexibility.  The experimental ingredients are, next to atoms, a static magnetic field, a modulation, and an RF field. The Zeeman levels of the atoms in the static trap are coupled by the RF-field resulting in dressed atomic potentials.  These potentials are subjected to the modulation. 
If the frequency $\omega_{\text{m}}$ of the modulation is fast compared to the
trap frequency of the static trap but still much smaller than the Larmor frequency then the atoms experience the time-average of the modulated dressed potential. An example of this can be seen in Fig.\ \ref{fig:TAAP}.  The conditions on the frequencies can be summarized as
\begin{equation}
\omega \ll \omega_{\text{m}} \ll \Omega_{\LA}.
\label{eq:FrequencyRegimes}
\end{equation}

TAAPs can be produced starting from any static magnetic
field configuration $\mathbf{B}(\mathbf{r})$, e.g.\ from a
Ioffe-Pritchard or a quadrupole field. The addition of the RF-dressing field
$\mathbf{B}_{\RF}(t)= \mathbf{B}_ {\RF}\sin \omega_ {\RF} t$ results in the adiabatic
potential \cite{Zobay2001PRL, Schmiedmayer1995PR, Lesanovsky2006PR}
\begin{equation}
V(\mathbf{r})=\mF \hbar \sqrt{\left(\Omega_{\LA}-
\omega_{\RF}\right)^2+\Omega_{\RF}^2} \label{eq:VIF}
\end{equation}
where  $\mF$ refers to the magnetic quantum number of the total atomic spin in the RF-dressed frame. $\Omega_{\RF}=|\gF \muB/ 2\hbar|\cdot 
\left|{\mathbf{B}(\mathbf{r})}/{|\mathbf{B}(\mathbf{r})|} \times
\mathbf{B}_{\RF}\right|$ is the RF coupling strength. The adiabatic potential
$V(\mathbf{r})$ is then modulated by periodically varying one or more of the quantities
$\Omega_{\LA}$, $\omega_{\RF}$, and $\Omega_{\RF}$ in accordance
with the inequality (\ref{eq:FrequencyRegimes}). 
This turns $V(\mathbf{r})$ into the time dependent potential $ V_{\mathrm{m}}(\mathbf{r},t)$ from which we can then calculate the TAAP by time-averaging over one period $\tau=2\pi\omega^{-1}_ {\text{m}}$ of the modulation:
\begin{equation}
V_{\TA}(\mathbf{r})=\frac{1}{\tau}\int_{0}^{\tau}V_{\mathrm{m}}(\mathbf{r},t)\,d
t. \label{eq:VTA}
\end{equation}

In order to illuminate the versatility of TAAPs we will now discuss
two examples: a ring and a double-well trap. The ring-shaped potential is much sought after as a waveguide in
atom interferometry \cite{Kasevich2002S}. It can also serve as a perfectly smooth duct in the study of persistent superfluid
currents \cite{Rokhsar1997PRL}.  The ring as a whole 
 is a Mexican-hat-shaped trap that can be used for the stable confinement of highly charged vortex and quantum Hall states \cite{Regnault2003PRL,
Kasamatsu2002PRA}. The double-well traps on the other hand have a tuneable barrier between the two parts of the trap.  Therefore, they are useful for tunnelling experiments  or as  ``beam-splitters'' for condensates. 

\begin{figure}[htbp]
\centering
\includegraphics[width=8cm]{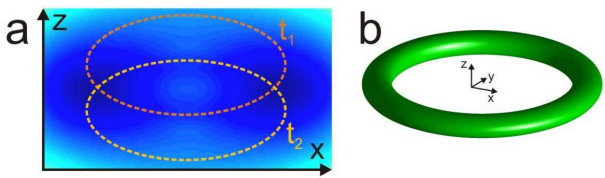}
\caption{ (color online) Generation of a ring-TAAP. (a) Density plot
of the TAAP in the $x z$-plane. The dashed lines sketch the
shell-like adiabatic RF potential at the times $t_{1}$ and $t_{2}$
where the modulation field has an opposite phase. (b) Representation
of a 3D isopotential surface of the ring-TAAP.} \label{fig:RingTAAP}
\end{figure}

Ring traps can be produced in a modulated magnetic quadrupole trap by RF-dressing the atomic levels. The vector of the static quadrupole field plus the modulation field
$\mathbf{B}_{\text{m}}(t)= \,B_
{\text{m}}\mathbf{e}_{\mathrm{z}}\sin \omega_ {\text{m}} t$ reads $\mathbf{B}(\mathbf{r},t)=
\alpha x\, \mathbf{e}_x+ \alpha y\,\mathbf{e}_y -(2 \alpha z-B_
{\text{m}} sin \omega_{\text{}m}t)\mathbf{e}_z$
 with $\alpha$ being the
gradient of the quadrupole. Adding the homogeneous RF-field
$\mathbf{B}_{\RF}(t)= \,B_ {\RF}\mathbf{e}_{z}\sin \omega_ {\RF}
t$ results in an adiabatic potential shaped like the shell of an
oblate spheroid (Eq.\,(\ref{eq:VIF})).  Note that $\mathbf{B}_{\text{m}}(t)$ simply displaces the adiabatic shell-like potential in the $z$-direction by  $|\mathbf{B}_ {\text{m}}(t)|/2\alpha$ (see Fig.\,\ref{fig:RingTAAP}).  By
modulating next to $\mathbf{B}_ {\text{m}}(t)$ also $\omega_{\RF}$ one can ensure that the dressed potential at all times assumes its minimum  on the ring $z_{\text{min}}=0$, $\rho_{\text{min}}=\rho_{0}$ with the radius $\rho=\sqrt{x^2+y^2}$. In order to achieve this the RF-frequency has to be 
$\omega_{\RF}(t)=\omega_{\RFC} \sqrt{1+\left(\beta \sin
\omega_{\text{m}}t\right)^{2}}\,$, where $\beta=\kappa B_{\text{m}}/\hbar
\omega_{\RF}$ is the 
modulation index  and $\omega_{\RFC}$ the carrier RF-frequency. The TAAP then also has a ring-shaped minimum at  $\rho_{0}=\hbar \omega_{\RFC}/ (\kappa \alpha)$ which can also be found directly using Eq.\,(\ref{eq:VTA}).  

We now calculate the trap frequencies of the ring TAAP.  Assuming $ \Omega_{\RF} \ll \omega_{\RF}$
allows us to disregard the spatial dependence of the
RF-coupling frequency $\Omega_{\RF}$ over the size of the trap
\footnote{The analytic expressions including the full spatial dependence of $\Omega_{\RF}(\mathbf{r})$ are rather unwieldy and
the correction to the trap frequencies are less than 0.5\% for
an apect ratio of 0.5 to 1.5 and $\Omega_{\RF} /
\omega_{\RF}<0.1$.}. However, we still account for the fact that the modulation $\mathbf{B}_{\text{m}}(t)$ globally shifts the quadrupole field which affects the
projection of the $\mathbf{B}_{\RF}$ onto $\mathbf{B}(\mathbf{r},t)$ and thus $\Omega_{\RF}$. 
The maximum trapping frequency  can be achieved by keeping $\Omega_{\RF}$ constant. This is because large trapping frequencies occur at a small $\Omega_{\RF}$ which results in trap losses due to Majorana spin flips. These losses are minimized at a constant $\Omega_{\RF}$, which can be achieved by   modulating the  RF
amplitude according
to $B_{\RF}(t)=B_{\RF\text{\tiny{0}}} \sqrt{1+\left(2 \beta \sin
\omega_{\text{m}}t\right)^{2}}$. From Eq.\,(\ref{eq:VTA}) we then obtain the trap frequencies
\begin{eqnarray}
\omega_{\uprho}&=& \omega_{0}\,\left( {1 + \beta^2 } \right)^{ - 1/4}\nonumber\\
\omega_{\mathrm{z}}&=&2\omega_{0}\sqrt{1-\left(1+\beta^2\right)^{-1/2}}\label{eq:RingTrapFrequency}
\end{eqnarray}
with $\omega_{0}=\frac{m_{\mathrm{F}} g_{\mathrm{F}} \mu_{\mathrm{B}}\,   
\alpha\, }{\left(m\,\hbar \Omega
_{\RF}\right)^{1/2}}$ and $m$ being the mass of the atom.  Equal trap frequencies in
the $z$- and $\rho$-direction occur at a modulation index of
$\beta=\nicefrac{3}{4}$  with
$\omega_{\mathrm{z}}=\omega_{\uprho}=2\omega_{0}/\sqrt{5}$. The
maximal confinement on the other hand, as defined by the average
trap frequency
$\bar{\omega}\equiv\sqrt{\omega_{\uprho}\omega_{\text{z}}}$, is
achieved at $\beta=\sqrt{3}$, where
$\omega_{\text{z}}=2\omega_{\uprho}=\sqrt{2}\,\omega_{0}$ and
$\bar{\omega}=\omega_{0}$. We note, that by changing the RF-frequency the diameter of the ring can be tuned from a few micrometers to a centimeter. If at the same time we keep $\beta$ constant then the radial confinement remains unaffected by the changes. This  is particularly advantageous for matter-wave interferometers as it allows the size, and thus sensitivity, greatly to be tuned without affecting any other experimental parameter. Experiments on persistent superfluid currents will profit from the fact that it is possible to introduce potential barriers in the rings. At $x=0$ this can be done by simply adding a small modulation field along $\mathbf{e}_{y}$.

\begin{figure}[htbp] 
\centering
\includegraphics[width=8cm]{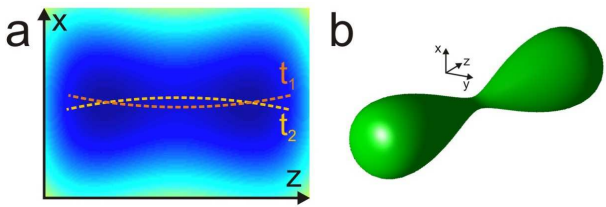}
\caption{ (color online) Generation of the double-well TAAP. (a)
Density plot of the TAAP in the $x z$-plane. The dashed lines sketch
the shell-like adiabatic RF potential at the times $t_{1}$ and
$t_{2}$ where the modulation field has an opposite phase. The
curvature is exaggerated. (b) A 3D representation of an isopotential
surface of the resulting TAAP. The surface is rotationally symmetric
around the $z$-axis around which the modulation field
$\mathbf{B}_{\text{m}}(t)$ rotates.} \label{fig:example}
\end{figure}

The second TAAP example we discuss is the double-well trap. The two wells can be
merged into a single potential minimum. Moreover, the individual traps can be
adiabatically transformed from a cigar into a
pancake shape. We start as in the ring case from a quadrupole field and add Fthe RF-field
$\mathbf{B}_{\RF}(t)$. For the double well trap we superimpose
a modulation field which rotates around the $z$-axis:
$\mathbf{B}_{\text{m}}(t)=B_{\text{m}} (\mathbf{e}_x \cos
\omega_{\text{m}} t + \mathbf{e}_y  \sin \omega_{\text{m}} t )$. This situation is shown in Fig.\,\ref{fig:example}.
If $0<\beta \ll 1$ and $kT<\hbar\Omega_{\RF}\ll \hbar\omega_{\RF}$ then we can
again disregard the spatial dependence of
$\Omega_{\RF}$ over the size of the atomic cloud. The resulting TAAP possesses two potential
minima located at $z_{0}= \pm \frac{\hbar \omega_{\RF}}{g_{\mathrm{F}} \mu_{\mathrm{B}}\, 2 \alpha}\sqrt{1-\beta^2}$. The two traps are identical and possess the
radial and vertical trap frequencies $\omega_{\uprho} =
\omega_0\,\frac{\beta}{\sqrt{2}}$ and $\omega_{\text{z}}= 2
\omega_0\ \sqrt{1-\beta^2}$, respectively
\footnote{The numeric
solution including the full spatial dependence of $\Omega_{\RF}(\mathbf{r})$  shows
that the error in $\omega_{\uprho}$ is smaller than 0.5\% for
$0<\beta<1$, and for $\omega_{\text{z}}$ smaller than 1\% for
$0<\beta<0.8$.}. As the value of $\beta$ is increased from $\beta\ll1$ to $\beta>1$ the trap transforms from pancake to cigar shape.  At $\beta=0.94$ the trap is spherical and  
at $\beta=1$ the two traps merge into a
single trap. Including the spatial dependence of $\Omega_{\RF}$, the
harmonic aspect ratio is then
$\omega_{\text{z}}/\omega_{\uprho}=2^{\nicefrac{3}{2}}B_{\RF}/B_{\text{m}}$. 
 It is interesting to note that, even though they have yet to be observed experimentally, these traps must always
exist during the RF-evaporation phase in standard TOP-traps, albeit that the atoms in the center of the TOP trap are in the anti-trapped state with respect to the double-well trap.\\

One of the very useful aspects of any TAAP is the ability to create multiple TAAP traps or waveguides simply by using multiple RF-frequencies instead of a single one.  These traps can either be fully separated or have tunable tunneling barriers between them. Multiple traps can be merged into a single one, or a single trap be split. If the difference between the RF-frequencies is much larger than the RF-coupling then the atoms interact at any one point of the TAAP only with a single sideband of the RF field \cite{Courteille2006JPB}. The resulting individual traps can then be calculated using Eq.\,(\ref{eq:VIF}) and (\ref{eq:VTA}).  However, in many cases, e.g.\ for tunneling experiments, one is interested in traps that are closely spaced, in which case multiple sidebands have to be taken into account at the same time. Double-wells, for example, require three RF-frequencies: one for the barrier and one for each of the wells referred to as the carrier and the two sidebands, respectively. This can conveniently  be realized by modulating the amplitude of the RF field: $B_{\RF}(t)= \,B_ {\RF}(\sin \omega_ {\text{M}}t)( \sin \omega_ {\RF} t)$.  The Fourier spectrum then consists of a carrier frequency $\omega_{\RF}$ and symmetrical two sidebands at $\omega_{\RF} \pm\omega_ {\text{M}}$. The carrier and sidebands result in the RF coupling  $\Omega_0$ and  $\Omega_1$, respectively.   Assuming $\Omega_1 \ll \omega_{\RF}$ one can write the resulting dressed double-well potential as
\begin{equation}
V(r)=\mF \hbar \sqrt{\left(\sqrt{\Omega^{2}_{\ell}+\Omega_{0}^{2} }-\omega_{\text{\tiny{M}}}\right)^{2}\!\! + \frac{\Omega_{1}^{2}}{1\!+\!\left(\Omega_{0}/\Omega_{\ell}\right)^{2}}} \label{eq:DoubleTrap}
\end{equation}
where $\Omega_{\ell}=\Omega_{\LA}(r)-\omega_{\RF}$. A similar approach has been used to investigate RF-evaporation in a dressed potential with a single minimum \cite{Alzar2006APP}. The smallest spacing of the two wells in Eq.\ (\ref{eq:DoubleTrap}) is for most experimental conditions is smaller than the harmonic oscillator length of the ground states of the individual traps 
\footnote{The distance $\Delta \rho=2 h \omega_{\text{M}}/(g_{\mathrm{F}} \mu_{\mathrm{B}} \alpha) $ of the two minima of the double well is limited by the requirement $\omega\ll\omega_{\text{\tiny{M}}}$ (see Eq.\ (\ref{eq:FrequencyRegimes})). Assuming $^{87}$Rb at a gradient of 5\,Tm$^{\text{\tiny{-1}}}$ and a trapping frequency in the individual well of 500\,Hz, the minimum spacing would be 90\,nm which is much shorter than the harmonic oscillator length of 480\,nm.}. Using Eq.\ (\ref{eq:VTA}) and (\ref{eq:DoubleTrap}) one can calculate the TAAP trap parameters for example for a double ring or double pancake.  Stacks of pancakes can be used to study coupled 2D quantum gases. Concentric ring arrays allow the construction of an array of  matter-wave interferometers.  This would considerably improve the sensitivity of the interferometer by improving the counting statistics since a large number of atoms spread over multiple traps could be used in a single measurement.  In contrast, 
 placing a large number of atoms in a single guide reduces the interference signal due to the nonlinear interaction of the atoms. Additionally, two coupled rings can be used to study spontaneous emergence of angular momentum Josephson oscillations  \cite{Lesanovsky2006XXX}. \\
 
 Decoherence is one of the crucial factors in any trapping scheme. It determines the transport properties  of a waveguide and limits the storage time of the trapped atoms.  Any corrugation in  the waveguide, for example, leads to a coupling of the longitudinal to the transverse modes, thus destroying the coherence of the atoms as they travel along the guide.  This severely limits the maximum length the atoms travel in the guide and thus reduces the sensitivity of the interferometer. Guides created from static magnetic fields have very elongated field generating structures. Imperfections in the field generating structures tend to corrugate the guiding potential and thus limit the maximum coherent transporting length to less than 150\,\micro\meter\ \cite{Wang2005PRL}.  As mentioned in the introduction, this problem does not occur in TAAP traps. Such traps and waveguides can be much smaller than the field generating coils and are also located far away from them.
The influence of corrugations in the field generation structure falls off exponentially with distance. Therefore they do not perturb the trapping potential.
 This will make it possible, for example, to create large area matter-wave  interferometers where the atoms need to be guided over large distances. Guided matter-wave  Sagnac interferometers would be excellent candidates for a detection of the Lense-Thirring effect.      
 Guided matter-wave interferometers in the Michelson configuration could be used in precision measurements of the Newtonian gravitational constant.

At low atom-densities the main limit of the trap lifetime are spin-flip losses
\cite{Majorana1932NC, Pritchard1983PRL}. An upper limit for the probability for a spin-flip transition to occur in a single passage through the potential minimum can be expressed in terms of the Landau-Zener parameter $\Gamma$\  as  $P=e^{-2\pi \Gamma} $
\cite{Zener1932PRS, Rubbmark1981PR}.  The Landau-Zener parameter itself can elegantly be written in
terms of the RF-coupling frequency $(\Omega_{\RF})$ and time derivative of the Larmor frequency at the center of the trap ($\Omega_{\LA}$) as $\Gamma=\Omega_{\RF}^2 \dot \Omega_{\LA}^{-1}$ \footnote{
The familiar Landau-Zener probability for an atom crossing the center of a Ioffe-Pritchard trap can be regained by setting $B_{\RF}=B_{0}$, and $\dot \Omega_{\LA}=\gF \muB \alpha\, v/\hbar $, where $v$ is the velocity of the atom and $\alpha$ as before the gradient.}. For $P\ll1$ the lifetime $\tau$ of atoms in such a trap is $ \tau\cong f^{-1}e^{2\pi \Gamma}$, where $f$ is the frequency with which the atom passes through the trap minimum.    We can now calculate an upper limit for the lifetime in the symmetric, ring-shaped TAAP trap
$(\omega_{\uprho}=\omega_{\text{z}})$, assuming $ k_{B}T\ll \hbar
\omega_{\RF}$ one
can locally neglect the curvature of the dressed potential shell. In
order to get an analytic upper limit for $\tau$, we take into account only
atoms which periodically pass through the RF-coupling region due to $\mathbf{B}_{\text{m}}(t)$.  Defining the
adiabaticity factor $A=\frac{\Omega_{\RF}}{\omega_{\text{m}}} \sqrt{\frac{\hbar \Omega{\RF}}{k
T}}$ one can write an upper limit of the loss rate $w$ as $w=\frac{ \sqrt{2}}{3
\pi}\omega _{\text{m}}^{2} A\ G_{0,3}^{3,0}\left(\left.\frac{2 \pi
^2}{9} A^2  \right|  \nicefrac{-1}{2},0,0 \right)$ where $G$ is the Meijer $G$-function.  From this we numerically find a lower limit for the lifetime to be $\tau=33\,\omega _{\text{m}}^{-2} e^{1.2A}$ in the experimentally interesting range $2<A<18$.

Let us now turn to the experimental ingredients needed: the
static quadrupole and the modulation and the RF fields.  We will base our numbers on $^{87}$Rb.  The main
confining gradient results from the static quadrupole which can be
constructed from a pair of anti-Helmholtz coils. A gradient of 4\,T\,m$^{-1}$ is readily achievable resulting in a maximum trap frequency of 1.3\,kHz for a 1\,\micro\kelvin\ cloud trapped with a lifetime of 60\,s. The homogeneous modulation fields must
oscillate fast compared to the static trap frequency, yet slow compared to the RF-coupling frequency.  Even for the above mentioned 1.3\,kHz trap, a modulation
frequency of only $\omega_{\text{m}}=2 \pi 13$\,\kilo\hertz\ suffices.  One can therefore use low-noise audio frequency generators together
with high-power audio amplifiers. Using a 4\,kW car amplifier
with matching coils one can reach modulation fields of 1\,mT.  Three
sets of coils then can offer complete control over the
 modulation fields.  The rather complex RF-modulation
needed are easily generated using an RF arbitrary waveform
generator.  A 20\,W  broad-band RF-amplifier coupled to a
non-resonant antenna suffices to reach the needed $\Omega_{\RF}$.
Note, that any change in the trapping configuration just requires a
change in frequencies and amplitudes of the RF- and modulation fields which can be computer controlled.

In conclusion, we have presented a novel class of trapping potentials in which a great
variety of different trap-shapes can be produced in a single
experimental apparatus. We have presented analytical expressions for
pancake, cigar, and ring shaped traps. The
different trap shapes can continuously be deformed from one into
another, by purely changing the RF- and modulation frequencies and
amplitudes. Multiple traps, e.g.\ stacks of pancakes or concentric
rings, can be formed. These traps can be merged or coupled through
well-controlled tunneling barriers. The coherence of a traveling matter-wave can be preserved over large distances due to the complete lack of corrugations in the TAAP potentials. This will make possible ultrahigh precision guided matter-wave interferometers.

This research project has been supported by a Marie 
Curie Transfer of Knowledge Fellowship (IL) and a Marie 
Curie Excellence Grant (WK) of the European Commu- 
nitys Sixth Framework Programme under the contract 
numbers MTKD-CT-2004-014496 and MEXT-CT-2005- 
024854.

\bibliography{TAAP}
\end{document}